\documentclass[prb,twocolumn,superscriptaddress,showpacs,aps]{revtex4}

\usepackage{color}
\usepackage{graphicx}
\usepackage{bm}
\usepackage{amsmath}

\begin{document}

\newcommand{\Ba}{BaFe$_2$As$_2$}
\newcommand{\BaCo}{Ba(Fe$_{0.93}$Co$_{0.07}$)$_2$As$_2$}
\newcommand{\Tm}{TmNi$_2$B$_2$C}
\newcommand{\Lu}{LuNi$_2$B$_2$C}
\newcommand{\Y}{YNi$_2$B$_2$C}
\newcommand{\Tc}{T_{\text{c}}}
\newcommand{\Hci}{H_{\text{c1}}}
\newcommand{\Hcii}{H_{\text{c2}}}
\newcommand{\fq}{\phi_0}
\newcommand{\lamn}{\lambda_n}
\newcommand{\TN}{T_{\text{N}}}

\title{Vortex studies in superconducting \BaCo}

\author{M.~R.~Eskildsen}
\email{eskildsen@nd.edu}
\affiliation{Department of Physics, University of Notre Dame, Notre Dame, IN 46556}

\author{L.~Ya.~Vinnikov}
\affiliation{Institute of Solid State Physics RAS, Chernogolovka, Moscow Region, 142432, Russia}

\author{T.~D.~Blasius}
\altaffiliation{REU student from University of Michigan.}
\affiliation{Department of Physics, University of Notre Dame, Notre Dame, IN 46556}

\author{I.~S.~Veshchunov}
\author{T.~M.~Artemova}
\affiliation{Institute of Solid State Physics RAS, Chernogolovka, Moscow Region, 142432, Russia}

\author{J.~M.~Densmore}
\affiliation{Department of Physics, University of Notre Dame, Notre Dame, IN 46556}

\author{C.~D.~Dewhurst}
\affiliation{Institut Laue-Langevin, 6 Rue Jules Horowitz, F-38042 Grenoble, France}

\author{N.~Ni}
\author{A.~Kreyssig}
\author{S.~L.~Bud'ko}
\author{P.~C.~Canfield}
\author{A.~I.~Goldman}
\affiliation{Ames Laboratory and Department of Physics and Astronomy, Iowa State University, Ames, Iowa 50011, USA}

\date{\today}

\begin{abstract}
We present small-angle neutron scattering (SANS) and Bitter decoration studies of the superconducting vortices in \BaCo. A
highly disordered vortex configuration is observed at all measured fields, and is attributed to strong pinning. This
conclusion is supported by the absence of a Meissner rim in decoration images obtained close to the sample edge. The field
dependence of the magnitude of the SANS scattering vector indicates vortex lattice domains of (distorted) hexagonal
symmetry, consistent with the decoration images which show primarily 6-fold coordinated vortex domains. An analysis of the
scattered intensity shows that this decreases much more rapidly than expected from estimates of the upper critical field,
consistent with the large degree of disorder.
\end{abstract}


\maketitle

The recent discovery of superconductivity in LaFeAsO,\cite{Kamihara08} with critical temperatures increasing to 38~K upon
doping or 43~K when subjected to hydrostatic pressure,\cite{Takahashi08,Chen08} has sparked a strong interst in these
materials. However, synthesis, and especially the growth of single crystals, of these materials have proven to be a
challenge. It was therefore a significant step forward when it was reported that the intermetallic compound \Ba \ could be
rendered superconducting by doping either K on the Ba-site,\cite{Rotter08,Ni08a} or Co on the Fe-site.\cite{Sefat08,Ni08b}
This discovery enabled the possibility of obtaining large, high-quality single crystals grown from flux.

Here we report on combined small-angle neutron scattering (SANS) and high resolution Bitter decoration studies of the
vortices in superconducting \BaCo.\cite{Ni08b} The SANS measurements show a ring of scattering, indicating a highly
disordered vortex configuration, and with a rocking curve extending beyond the measurable range. This result is confirmed
and extended to lower fields by the Bitter decoration images. The disordering is attributed to strong vortex pinning, which
is supported by the absence of a Meissner rim in decoration images obtained close to the sample edge. The field dependence
of the magnitude of the SANS scattering vector indicates vortex lattice domains of (distorted) hexagonal symmetry. An
analysis of the scattered intensity due to the vortices shows a rapid decrease with increasing applied magnetic field,
significantly exceeding what would be expected based on estimates of the upper critical field.

Small-angle neutron scattering (SANS) measurements were performed on a 260~mg single crystal of \BaCo. The single crystal
used in the SANS experiment was grown from a FeAs/CoAs mixture, and had a critical temperature $\Tc = 21$~K with a
relatively narrow transition width $\Delta \Tc = 2$~K measured by dc magnetization.\cite{Ni08b}
The experiment was performed at the D22 SANS instrument at the Institut Laue-Langevin.  Incident neutrons with wavelength
$\lamn = 6 - 14$ {\AA} and wavelength spread of $\Delta\lamn/\lamn = 10\%$ were used, and the diffraction pattern
collected by a position sensitive detector.
Measurements were performed at 2~K in horizontal magnetic fields up to 810~mT, applied at angle of $5 - 10^{\circ}$ to the
$c$ axis to reduce background scattering from crystallographic defects. All vortex measurements were done at 2~K, following
a field cooling procedure. Background measurements obtained at 25~K $> \Tc$ were subtracted from the data.

The high resolution Bitter decoration experiments were performed on a \BaCo \ single crystal, on an as grown surface
perpendicular to the $c$ axis, for applied magnetic fields between 1 and 32~mT along the $c$ axis and following a field
cooling procedure. The crystal surface was flat and shiny except in areas with growth steps. The initial sample temperature
was $1.7$~K, increasing to 5 - 6~K at the end of the decoration process. The sample was subsequently imaged using a
scanning electron microscope (SEM) in the second electron emission regime, to locate the islands of iron particles which
had decorated the sample at the vortex positions.\cite{Vinnikov93}

\begin{figure*}
  \includegraphics{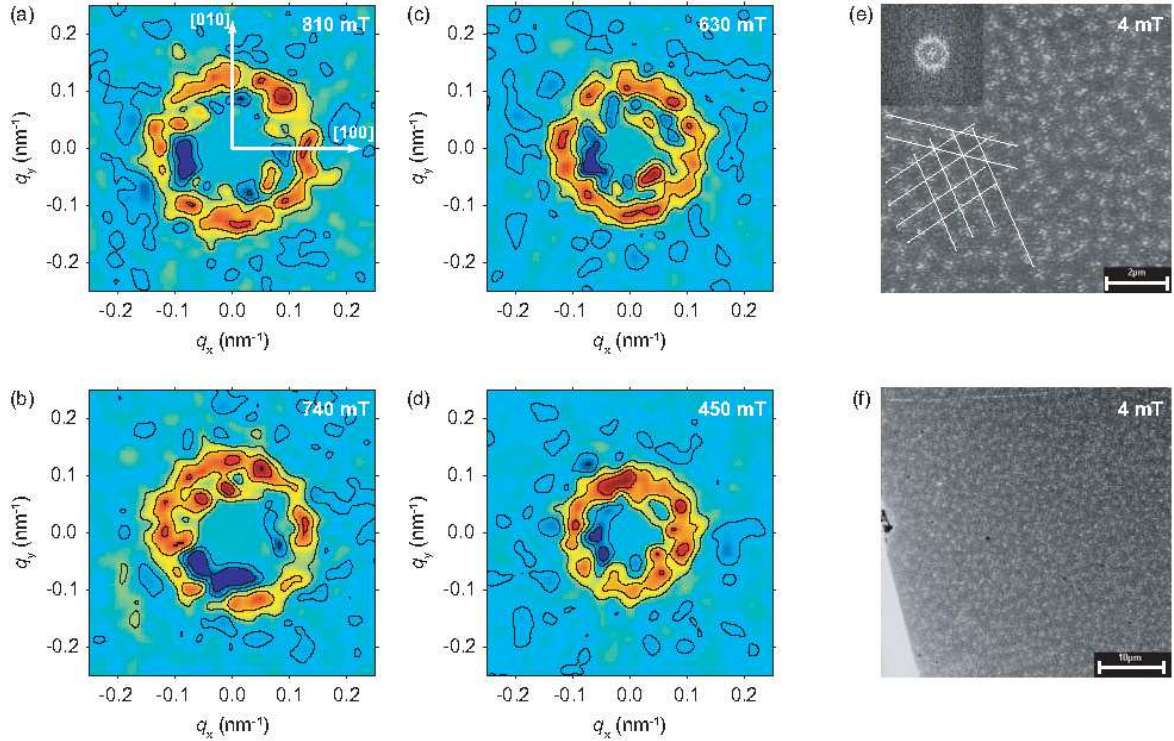}
  \caption{(Color online) Vortex imaging in \BaCo \ by SANS (a-d) and Bitter decoration (e, f).
           The SANS vortex diffraction patterns were obtained at 2~K and applied fields of 810 (a), 740 (b), 630 (c) and
           450~mT (d), following a field cooling procedure. The field was applied at a $5 - 10^{\circ}$ angle with respect
           to the crystalline $c$ axis to reduce background scattering from crystallographic defects.
           For all four fields data was collected at a single (centered) orientation of the magnetic field with respect to
           the neutron beam direction, and background measurements obtained at 25~K $> \Tc$ were subtracted. The data is
           smoothed and the central part of the detector is masked off.
           The decoration images were obtained in a magnetic field of 4~mT in the center of the sample (e), and in the
           vicinity of the sample edge (f). Small ordered domains can be observed as indicated in (e). The inset to (e)
           show the Fourier transform of the entire decoration image.
           \label{DifPatDecs}}
\end{figure*}

Fig.~\ref{DifPatDecs}(a-d) shows vortex diffraction patterns obtained at four different applied magnetic fields between
810 and 450~mT. 
For all the magnetic fields investigated, a disordered vortex ``lattice'' (VL) is observed, evident by a ``powder
diffraction'' ring of scattering instead of well defined Bragg peaks. It is reasonable to suppose that the disordering of
the VL is caused by pinning of vortices to defects in the sample.
Nonetheless, the ring of scattering indicates that short range order persists, with small
ordered VL domains randomly oriented with respect to one another. The intensity variation around the rings of scattering
suggest the existence of maxima along the horizontal and vertical axis corresponding to the \{100\} directions, indicating
a VL preferred orientation even if it insufficient to cause the coalignment of all domains.

Scattering from the vortices is expected at a radial distance from the center of the detector which is proportional to the
VL scattering vector, $q$. As the magnetic field, and hence the vortex density decreases, the average intervortex spacing
increases as $a \propto (\fq/B)^{1/2}$, where $\fq = h/2e = 20.7 \times 10^4 \mbox{ T\AA}^2$ is the flux quantum. Lowering
the field we therefore expect the ring of scattering to shrink, since $q \propto 2\pi/a \propto (B/\fq)^{1/2}$ decreases.
This is clearly seen in Fig.~\ref{DifPatDecs}(a-d), confirming that the scattering is indeed due to the superconducting
vortices.

Results of Bitter decoration measurements obtained in a magnetic field 4~mT are shown in Fig.~\ref{DifPatDecs}(e, f).  A
close inspection of the decoration image in Fig.~\ref{DifPatDecs}(e) reveals primarily 6-fold coordinated vortices, as well
as the existence of small, well ordered domains with a size of $\sim 5$ vortex spacings. However, no long-range
orientational order is established, as evident from the Fourier transform of the decoration pattern image shown in the
insert.
The Fourier transform is directly comparable to the diffraction patterns in Fig.~\ref{DifPatDecs}(a-d). Combined with the
SANS results this confirms that the disorder is predominantly orientational and longitudenal (along the vortices/field
direction), but not $\Delta d$-disorder (vortex spacing). Similar results were obtained at fields as low as 1~mT and as
high as 32~mT. Computing the average area associated with each vortex in a 1~mT decoration image (not shown) we find a
magnetic flux per vortex of $(21 \pm 1) \times 10^4 \mbox{ T\AA}^2$, in agreement with $\fq$. Further proof of strong bulk
pinning is the absence of a Meissner rim (low vortex density region) in the decoration image shown in
Fig.~\ref{DifPatDecs}(f), which was obtained near the edge of the sample.\cite{Vinnikov98}

Our observation of a disordered vortex arrangement is consistent with recent reports of STS imaging in \BaCo.\cite{Yin08}
However, compared to small-scale direct space imaging like STS, our results demonstrate that the disordering is a bulk
effect, not restricted to the sample surface or isolated regions.

\begin{figure}
  \includegraphics{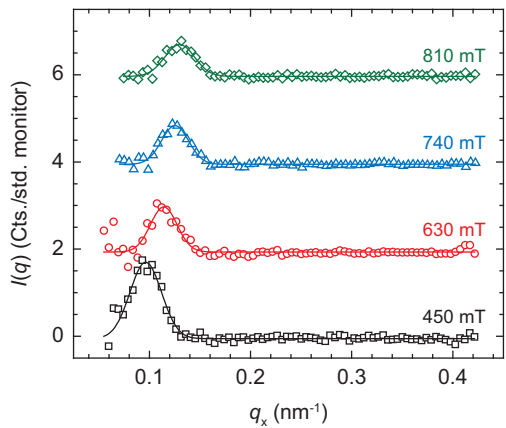}
  \caption{(Color online) Radial intensity distribution for the four diffraction patterns in Fig.~\ref{DifPatDecs}(a-d).
           The curves have been offset by 2~Cts./Std. Mon. for clarity. Also shown is a Gaussian fit to the data at each
           field.
           \label{IvsQ}}
\end{figure}

\begin{figure}
  \includegraphics{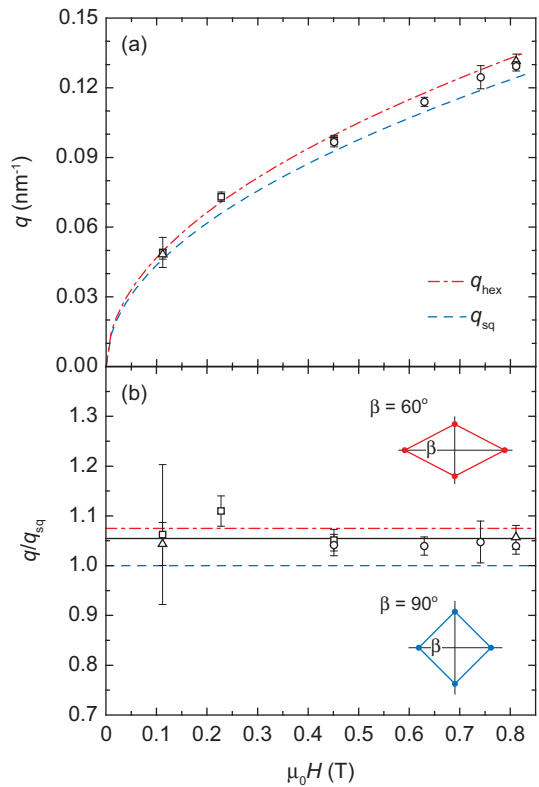}
  \caption{(Color online) Comparison of the measured scattering vector (a), and the ratio $q/q_{\text{sq}}$ (b), to that
           expected for respectively a square (dashed lines) and a hexagonal (dot-dashed lines) vortex lattice.
           The solid line in panel (b) is the average $q/q_{\text{sq}} = 1.055$.
           The different symbols corresponds to measurements performed at different neutron wavelength, $\lamn$.
           \label{QvsH}}
\end{figure}

The field dependence of the scattering vector is also evident from Fig.~\ref{IvsQ}, which show the radial intensity
distribution corresponding to each of the diffraction patterns in Fig.~\ref{DifPatDecs}(a-d).
Fitting each radial intensity curve to a Gaussian it is possible to obtain a quantatitive measure of both the magnitude of
the scattering vector and the scattered intensity. Furthermore the FWHM of the fits yield $\Delta q/q \approx 0.3$, which
is $\sim 50$\% larger than the experimental resolution and consistent with scattering from small ordered domains of size
$\sim (\pi \Delta q/q)^{-1}$, corresponding to a few VL spacings in agreement with the decoration results.
If one considers a rhombic VL unit cell as shown in the insets to
Fig.~\ref{QvsH}(b), the scattering vector is given by $q = 2 \pi (B/\fq \sin \beta)^{1/2}$ where $\beta$ is the opening
angle. Here $\beta = 60^{\circ}$ (or $120^{\circ}$) corresponds to a hexagonal VL and $\beta = 90^{\circ}$ to a square VL.
Fig.~\ref{QvsH}(a) shows the measured scattering vector as a function of applied magnetic field, compared to that expected
for respectively a square and a hexagonal VL.
With the exception of a single data point, all the measured scattering vectors fall between $q_{\text{hex}}$ and
$q_{\text{sq}}$, although it is closer to the former. This is consistent with the observation of primarily 6-fold
coordinated vortices in the decoration images. 
Fig.~\ref{QvsH}(b) shows the
measured scattering vector normalized to $q_{\text{sq}}$. Within the error bars, no field dependence of the normalized
scattering vector is observed over the measured field range.

While a hexagonal VL is expected for an isotropic material, it is also well known that most tetragonal superconductors
posses enough anisotropy to stabilise a square VL above a certain magnetic
field.\cite{Eskildsen97a,Vinnikov01,Eskildsen03,Brown04,Yethiraj05} The origin of the anisotropy can arise, either from the
Fermi surface of a given material coupled with non-local electrodynamics,\cite{Kogan97} or it can be due to anisotropic
pairing in the superconducting state.\cite{Ichioka99} Given that a near hexagonal VL symmetry is observed in \BaCo \ at all
the measured fields, it is tempting to conclude that the supercondcuting state in the basal plane of this material is
highly isotropic. However, it has been shown that, at least in the case of a square VL stabilised by a Fermi surface
anisotropy, the critical field for the transition to the square symmetry depends strongly on the so-called non-locality
range, which decreases with increasing levels of impurities (decreasing mean free path).\cite{Gammel99} From our results
it is not possible to conclude which mechanism is responsible for the hexagonal VL in \BaCo.

We now turn to the integrated scattered intensity, obtained from Gaussian fits to the radial intensity distribution as
shown in Fig.~\ref{IvsQ}. From a measurement of the absolute reflectivity, $R$, it is possible to determine the VL
form factor by
\begin{equation}
\left| F(q) \right| ^2 =  \frac{16 \fq^2 q}{2 \pi \gamma^2 \lamn^2 t} \; R,
\end{equation}
where $\gamma = 1.91$ is the neutron gyromagnetic ratio and $t$ is the sample thickness.\cite{Christen77}
However, to put the form factor on an absolute scale, the scattered intensity from the vortices should be integrated as
the sample and magnetic field is rotated with respect to the incoming neutron beam, causing the reflections to cut through
the Ewald sphere. In the case of \BaCo \ scattering from the vortices showed no measurable change in intensity even for
rotations of several degrees. This indicates that the high degree of disorder found in the basal plane is also present
along the direction of the vortices, causing them to deviate substantially from the applied field direction. In this
regard the results in \BaCo \ are reminiscent of those obtained on CaC$_6$.\cite{Cubitt07} Our results are consistent with
the report of Prozorov {\em et al.} of a highly disordered vortex phase in \BaCo, based on the analysis of transport
measurements.\cite{Prozorov08}

\begin{figure}
  \includegraphics{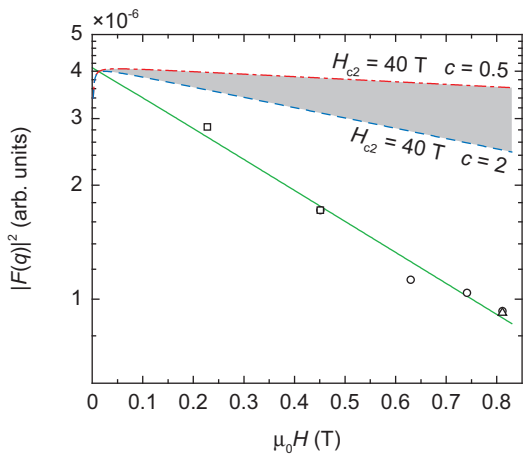}
  \caption{(Color online) Field dependence of the vortex scattered intensity. The different symbols corresponds to
           measurements performed at different neutron wavelength, $\lamn$.
           The solid line is an exponential fit to the data with a slope of $-0.82 \; (1/\mbox{T})$.
           The dashed and dot-dashed lines show the form factor calculated using the London model (Eq.~(\ref{FFLondon}))
           with $c = 2$ and and $1/2$, and $\Hcii = 40$~T.
           \label{FF}}
\end{figure}

Our inability to measure a proper rocking curve means that a determination of the form factor on an absolute scale is not
possible. On the other hand, if we assume a constant rocking curve width, the radial integrated intensity will be
proportional to $|F(q)|^2$. Due to the strong pinning in the sample, and relatively small field range investigated, we
we believe that this is a resonable assumption.
The form factors obtained in this fashion are shown in Fig.~\ref{FF} on a logarithmic scale.
Here it is important to stress that even though the absolute magnitude of $|F(q)|^2$ is not known, the relative field
dependence can still be determined. The data are well fitted by an exponential field dependence as shown by the solid line
in Fig.~\ref{FF}.

Several models exist for the form factor field dependence. By far the simplest is based on the London model, extended by
a Gaussian cut-off to take into account the finite extent of the vortex cores:
\begin{equation}
  F(q) = \frac{B}{1 + (\lambda q)^2} \; e^{-c (\xi q)^2}.
  \label{FFLondon}
\end{equation}
Here $\lambda$ and $\xi$ are respectively the penetration depth and coherence length, and the constant $c$ is typically
taken to be between $1/4$ and 2.\cite{Yaouanc97} Even though more realistic models for the form factor are available, the
London model is deemed adequate for the following discussion since the SANS measurements were performed at
$T \approx 0.1 \, \Tc$ and $H \ll \Hcii$. For simplicity we will also assume a square vortex lattice for the following
analysis. It should be stressed that neither of these choices affect the following discussion in any significant fashion.
For all fields studied $(q\lambda)^2 \gg 1$ and since $q^2 = (2\pi)^2 B/\fq$ the form factor in Eq.~(\ref{FFLondon})
decreases exponentially with increasing magnetic field.

As shown in Fig.~\ref{FF}, the measured intensity is indeed well fitted by an exponential decrease. If we use the fitted
slope to estimate the superconducting coherence length we obtain $\xi = 10$~nm. Here we have used $c = 0.5$ which in the
past been found to yield reasonable values for $\xi$ in other materials.\cite{Eskildsen97b} However, this value of  $\xi$
is roughly $3.5$ times higher than expected from the extrapolated upper critical field of $\approx 40$~T at
2~K.\cite{Ni08b} For comparison the $\Hcii$ corresponding to this ``measured'' $\xi$ is only $3.3$~T, more than an order of
magnitude less than the measured upper critical field. The difference is emphasized in Fig.~\ref{FF}, which shows the
range of the calculated form factor (shaded area) corresponding to the real $\Hcii$ and the two extreme values of $c$. From
this it is clear that the measured form factor falls off much faster than expected. We believe that this rapid decrease in
the scattered intensity is due to the high degree of disorder of the ``vortex lattice'' in \BaCo. This result is indicative
of the existence of a vortex glass or randomly oriented domains of a Bragg glass, consistent with the results of Klein
{\em et al.}\cite{Klein01}

In summary, we have performed SANS and Bitter decoration studies of the vortices in superconducting \BaCo. At all the
fields investigated, the results show a highly disordered vortex configuration was observed, most likely due to strong
pinning in this material. Further measurements at higher fields and temperatures should be carried out, to fully explore
the vortex phase diagram in this and related comppounds.

This work is supported by the National Science Foundation through grant DMR-0804887 (M.R.E, J.M.D.) and
PHY-0552843 (T.D.B.).
L.Y.V. and I.S.V. thank grant RFBR 07-02-00174 for support.
Work at the Ames Laboratory was supported by the Department of Energy, Basic Energy Sciences under Contract No.
DE-AC02-07CH11358.


\newpage

\printfigures

\end{document}